\documentclass[12pt,apj]{emulateapj}
\usepackage{graphicx}
\usepackage{subfigure}
\usepackage[usenames]{color}

\begin{document}

\shorttitle{Impact of type II spicules in the corona}
\shortauthors{Mart\'inez-Sykora et al.}
\title{Impact of type II spicules in the corona: Simulations and synthetic observables}

\author{Juan Mart\'inez-Sykora \altaffilmark{1,2} \& Bart De Pontieu \altaffilmark{2,3,4} 
\& Ineke De Moortel\altaffilmark{5} \& Viggo H. Hansteen \altaffilmark{3,4} \& Mats Carlsson \altaffilmark{3,4} }
\email{juanms@lmsal.com}
\affil{\altaffilmark{1} Bay Area Environmental Research Institute, NASA Research Park, Moffett Field, CA  , CA 94952, USA}
\affil{\altaffilmark{2} Lockheed Martin Solar and Astrophysics Laboratory, Palo Alto, CA 94304, USA}
\affil{\altaffilmark{3} Institute of Theoretical Astrophysics, University of Oslo, P.O. Box 1029 Blindern, N-0315 Oslo, Norway}
\affil{\altaffilmark{4} Rosseland Centre for Solar Physics, University of Oslo, P.O. Box 1029 Blindern, N-0315 Oslo, Norway}
\affil{\altaffilmark{5}School of Mathematics and Statistics, University of St Andrews, St Andrews, Fife, KY16 9SS, UK}

\newcommand{\newch}[1]{\color{red}{#1}}

\newcommand{\eg}{{\it e.g.,}} 
\newcommand{\myemail}{juanms@lmsal.com}
\newcommand{\komment}[1]{\texttt{#1}}
\newcommand{\ul}{\underline}
\newcommand{\pref}{\protect\ref}
\newcommand{\soho}{{\em SOHO{}}}
\newcommand{\sdo}{{\em SDO{}}}
\newcommand{\stereo}{{\em STEREO{}}}
\newcommand{\iris}{{\em IRIS{}}}
\newcommand{\hinode}{{\em Hinode{}}}
\newcommand{\B}{$\bullet$ }
\newcommand{\si}{\ion{Si}{4}~1402\AA }
\newcommand{\sir}{\ion{Si}{4}}
\newcommand{\he}{\ion{He}{2}~256\AA }
\newcommand{\her}{\ion{He}{2}}
\newcommand{\ca}{\ion{C}{4}~1548\AA }
\newcommand{\car}{\ion{C}{4}}
\newcommand{\fen}{\ion{Fe}{10}~184\AA }
\newcommand{\fern}{\ion{Fe}{10}}
\newcommand{\fex}{\ion{Fe}{10}~184\AA }
\newcommand{\ferx}{\ion{Fe}{10}}
\newcommand{\fet}{\ion{Fe}{12}~195\AA }
\newcommand{\fert}{\ion{Fe}{12}}
\newcommand{\fetsdo}{\ion{Fe}{12}~193\AA }
\newcommand{\fef}{\ion{Fe}{14}~211\AA }
\newcommand{\ferf}{\ion{Fe}{14}}

\begin{abstract}

The role of type II spicules in the corona has been a much debated topic in recent years. This paper aims to shed light on the impact of type II spicules in the corona using novel 2.5D radiative MHD simulations including ion-neutral interaction 
effects with the Bifrost code. 
We find that 
the formation of simulated type II spicules, driven 
by the release of magnetic tension, impacts the corona in various manners. Associated with the 
formation of spicules, the corona exhibits 1) magneto-acoustic shocks and flows  
which supply mass to coronal loops, and 2) transversal magnetic  
waves and electric currents that propagate at Alfv\'en speeds. The transversal waves 
and electric currents, generated by the spicule's driver and lasting for many minutes, are dissipated and heat the associated loop. These complex interactions in the corona can be connected with 
blue shifted secondary components in coronal spectral lines (Red-Blue asymmetries) observed with \hinode/EIS and \soho/SUMER, as well as the EUV counterpart of type II spicules and propagating coronal disturbances (PCDs) observed with the 171~\AA\ and 193~\AA\ \sdo/AIA channels. 

\end{abstract}


\section{Introduction}

Highly dynamic chromospheric jets that permeate the corona are one of the 
most common features observed at the solar limb. Consequently, given their occurrence rate, they have been suggested as contributors to coronal 
heating and/or as possible drivers of the solar wind \citep[e.g.,][]{Beckers:1968qe,Athay:1982fk,McIntosh:2011fk,De-Pontieu:2011lr,De-Pontieu:2017pcd}. When seen at the limb such jets are named spicules. Two types have been observationally identified \citep{de-Pontieu:2007kl,Tsiropoula:2012yq,Pereira:2014eu}: type I spicules are observed to have relatively slow rise velocities and small heights, type II spicules appear to rise much more rapidly and often extend to heights of at least $>10$~Mm. 
However, since the exact nature of type II spicules has been unknown, their mass and energy contribution to the corona has so far not been properly addressed. Hence, whether (and how much) type II spicules contribute or impact the corona and/or solar wind is under ongoing, vigorous debate 
\citep[e.g. compare with ][]{Madjarska:2011fk,Klimchuk:2012kx}.

Spicules are an observational phenomenon in which many different complex physical processes appear to be at work. In the current paper, we take a broad approach to the problem of understanding “coronal heating associated with spicules”. We do not limit ourselves to a simplifying scenario in which ad-hoc advection of expanding gas along a single field line dominates the energetics of spicules \citep{Klimchuk:2012kx,Klimchuk:2014fk}. Instead, our research takes a broader approach by studying the coronal heating associated with self-consistently generated features that show many similarities with observed properties of spicules by using a 2.5D radiative MHD numerical model. This model includes ambipolar diffusion which is a key process for the formation of type II spicules \citep{Martinez-Sykora:2017sci}. We thus continue the solar physics community’s quest to understand how much coronal heating is associated with spicules \citep[see also, e.g.,][and references therein]{Tsiropoula:2012yq,Pereira:2014eu}. 
Some known observational constraints and properties of type II spicules are 
listed as follows. These jets reach velocities of [40-150]~km~s$^{-1}$ and heights of 
$\sim10$~Mm prior to falling back some 3 to 10 minutes after their first 
appearance \citep{Pereira:2014eu,Skogsrud:2015qq}.  In contrast to 
type I spicules, some of the ejected chromospheric plasma of type II spicules 
is often heated to transition region (TR) temperatures. Therefore, 
 these spicules observed in the chromospheric \ion{Ca}{2} H line, e.g. with the Solar Optical Telescope on board of \hinode\ \citep[SOT,][]{Tsuneta:2008kc}, are typically short lived (1-2 minutes), whereas type II spicules observed in 
hotter or more opaque chromospheric lines such as \ion{Mg}{2} or TR lines such as \sir\, observed with \iris\ \citep{De-Pontieu:2014yu}, are longer lived (3-10 minutes). 

Type II spicules have been associated with various TR and 
coronal observables. Their on-disk counterparts, so called Rapid Blue-shifted Events (RBEs),
are associated with brightenings in EUV spectral lines that are sensitive to plasma at transition region and coronal temperatures \citep{De-Pontieu:2011lr} as observed with the Atmospheric 
Imaging Assembly \citep[AIA,][]{Lemen:2012uq} on board of the Solar Dynamic
Observatory (SDO) \citep[but consider also][]{Madjarska:2011fk}. 
Transition region and coronal spectral lines observed by the Extreme-ultraviolet Imaging Spectrometer \citep[EIS, ][]{Culhane:2007fk} onboard Hinode 
show blue shifted secondary components that could be associated with type II spicule dynamics \citep[e.g.][]{De-Pontieu:2009fk,McIntosh:2009lr,Peter:2010fk,Bryans:2010lr}. 
These features have nevertheless not been conclusively connected to type II spicules. 
More recently, Propagating Coronal Disturbances (PCDs) have been linked to 
type II spicules \citep{De-Pontieu:2017pcd}. Their observations show that coronal loops are formed and exist for many tens of minutes in association with spicules. To interpret these observations, recent models of type II spicules \citep{Martinez-Sykora:2017sci} were used. This radiative MHD model, based on the Bifrost code  \citep{Gudiksen:2011qy}, self-consistently generates jets that resemble type II spicules. In the current paper, we study in greater detail the impact of these simulated type II spicules on the corona.  Our results show that the complexity of our 2.5D MHD model invalidates many of the simplifying assumptions made in previous modeling attempts to address the coronal impact of type II spicules \citep[e.g. compare the results of this paper with][]{Klimchuk:2014fk}. 

Type II spicules are also associated with various types of waves which could, in principle, also contribute
to the driving of the solar wind and heating of the corona. Spicules show torsional \citep[e.g.][]{De-Pontieu:2012fu,De-Pontieu:2014fv} and
both low and high frequency transversal motions \citep{De-Pontieu:2007bd,Okamoto:2011kx,Srivastava:2017cl}. These studies provide estimates of the wave energy associated with spicules and it is claimed that waves carry enough energy flux into the corona to power the solar wind \citep{De-Pontieu:2007bd,McIntosh:2011fk}. However, it is unclear how this wave energy is generated and/or dissipated. Until now, we have not had access to realistic models for the generation of magnetic waves in spicules. In this paper we use the \citet{Martinez-Sykora:2017sci} model to shed some light on these issues.

Our paper starts with a description of the numerical models (Section~\ref{sec:sim}). Section~\ref{sec:res} details our results where we first describe transition region and coronal synthetic observables associated with the 
type II spicules (Section~\ref{sec:obs}), and determine the width of the coronal loops formed in association with the spicules (Section~\ref{sec:width}). We also analyze the flows and shocks that penetrate into the corona (Section~\ref{sec:flow}) as well as the mass and energy deposited into the corona (Section~\ref{sec:heat}). We  investigate the causes for the widths of the spicule-associated coronal loops and also the heating distribution along the loops (Section~\ref{sec:width2} and Section~\ref{sec:strat}). We finish with the discussion and conclusions (Section~\ref{sec:dis}).

\section{Simulation}~\label{sec:sim}

We use two types of 2.5D radiative MHD numerical simulations, both based on the Bifrost code \citep[see][for details on the numerical implementation and validation]{Gudiksen:2011qy}. The numerical methods implemented in our simulations include radiative transfer with scattering \citep{Skartlien2000,Hayek:2010ac,Carlsson:2012uq}, and thermal conduction along the magnetic field. The numerical domain in both simulations is identical and covers a region from the upper layers of the convection zone (down to $3$~Mm below the photosphere) into the corona (up to $40$~Mm above the photosphere). The spatial resolution along the vertical axis is non-uniform and ranges from 12~km (from the convection zone to the transition region) slowly increasing with height to 69~km (in the corona). The domain extends $96$~Mm in the horizontal direction with a spatial resolution of $14$~km. The energy flux entering the bottom of the computational domain and the chemical composition of the model is set to solar values. The only free parameter in these simulations is the initial distribution of magnetic field. In both cases, there is no new magnetic flux injected at the lower boundary and we have seeded two main opposite polarities of plage or enhanced network with unsigned magnetic field of 190~G. These regions are connected by $\sim 50$~Mm long loops. In particular, in both of these models, the hot corona is self-consistently maintained, despite the constraints inherent in 2.5D dynamics. Previous 2D simulations with a smaller numerical domain required the addition of a `hot plate' at the top boundary to maintain coronal temperatures due to their smaller numerical domain and/or lack of ability to resolve various small scale processes \citep{Martinez-Sykora:2017gol}.

\begin{figure*}[tbh]
	\begin{center}
		\includegraphics[width=0.95\hsize]{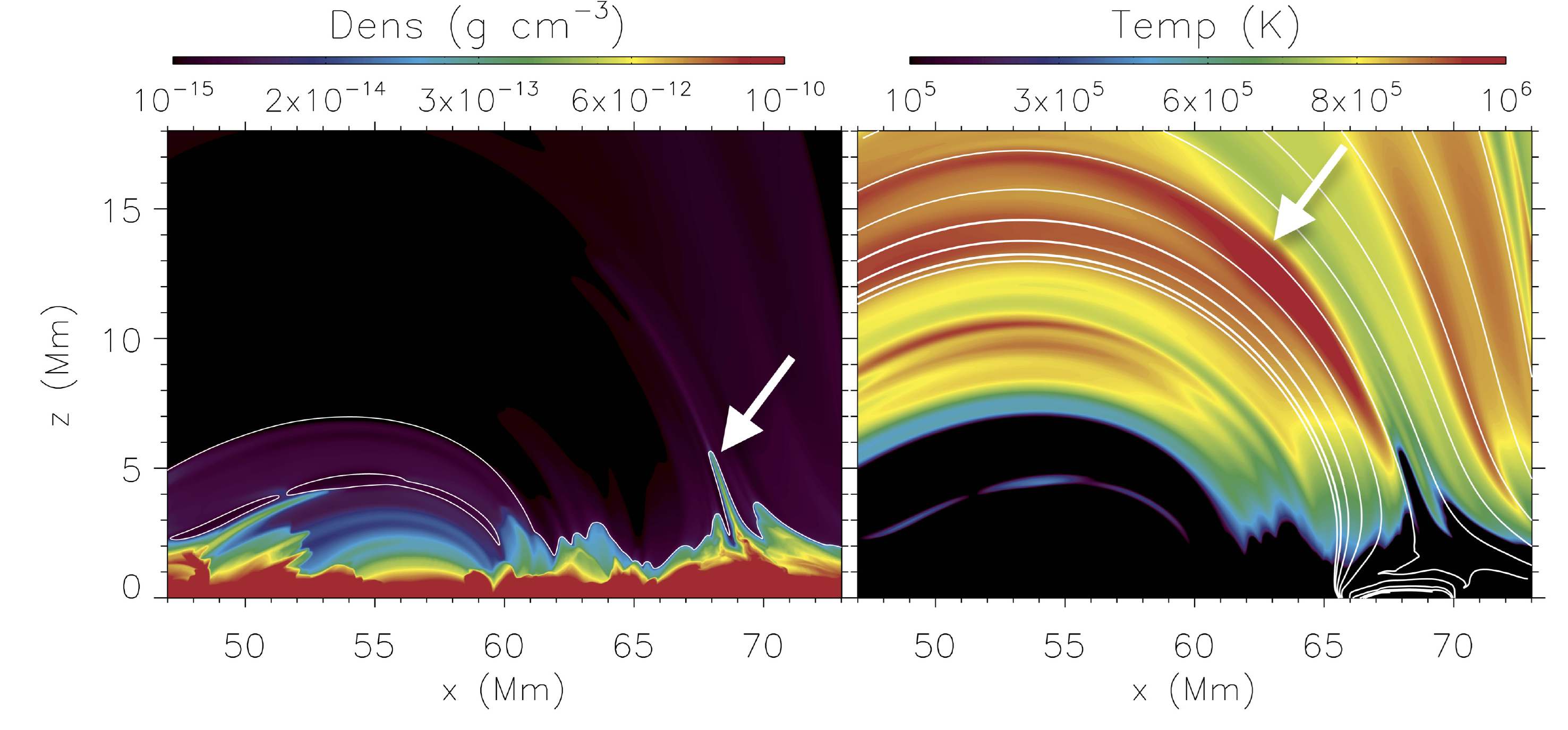}
		\caption{Density (left) and temperature (right) maps of the region of interest of the GOL simulation at $t=1270$~s. The spicule is marked with the arrow in the left panel and the loop associated with the spicule with the arrow in the right panel. Temperature at $T=10^5$~K are shown with white contours in the left panel. Selected field lines are shown in the right panel in white.} 
		\label{fig:init}
	\end{center}
\end{figure*}

In the ``GOL'' (Generalized Ohm's Law) simulation, we have included the effects of interactions between ions and neutrals in the partially ionized chromosphere by including ambipolar diffusion through using a generalized Ohm's law in the induction equation \citep{cowling1957,Braginskii:1965ul}. In this simulation, jets that resemble type II spicules are formed ubiquitously, as a result of ambipolar diffusion (e.g., see the example shown with the arrow in the left panel in Figure~\ref{fig:init}). In contrast, simulations without ambipolar diffusion, i.e., called the ``non-GOL'' simulation in \citet{Martinez-Sykora:2017gol}, as well as previous models \citep{Martinez-Sykora:2013ys,Carlsson:2016rt}, typically do not produce type II spicules, or if at all, only rarely.
Unless otherwise noted, we will refer to the GOL simulation when describing the impact of type II spicules on the corona. 
Figure~\ref{fig:init} shows that the chromospheric jet and associated hot loop are not always perfectly aligned along the same field lines. This is because the various contributing mechanisms for the heating and flows  (i.e., currents, waves, flows, and fluxes) do not always occur along exactly the same magnetic field line. Instead they occur on a range of magnetic field lines that either pass through or occur in close vicinity of the chromospheric spicule. These fields lines are all coupled together and connected to the driving mechanism that generates chromospheric spicules. The varying spatial relationship between the currents and flows is caused by variations in which field lines release tension, and which field lines carry currents (see also Martinez-Sykora et al 2017). The spatial offsets between the various contributing mechanisms are quite small and difficult to resolve at the spatial resolution of current coronal instruments ($>1$~arcsec).

We briefly summarize the main relevant results on the generation of type II spicules \citep[see][for details]{Martinez-Sykora:2017sci}: Simulated type II spicules are driven by at least, two different processes. 1) The release of magnetic energy, which has been built up by convective motions near magnetic flux concentrations, such as in network or plage regions. In such regions, the convective motions can impose substantial tension on the magnetic field. This tension often ``diffuses" away from regions where the gas pressure dominates (i.e., where plasma beta, $\beta >1$), towards regions where $\beta \approx1$, higher in the atmosphere. Ambipolar diffusion is the process that helps to move this magnetic field through the photosphere and lower chromosphere. At the same time, ambipolar diffusion helps to spatially confine the magnetic tension to thin layers. When this magnetic field eventually reaches heights where $\beta \approx 1$, the magnetic tension is released in a similar manner as the whiplash effect which gives rise to pressure gradients and thereby accelerates plasma. 2) Magneto convective motions perturb overlying and highly bent magnetic field producing magneto-acoustic waves traveling almost parallel to the surface. Once the shock reaches the highly bent magnetic field at heights where $\beta \approx 1$, the wave mode changes, and in combination with the release of some of the tension from the highly bent field lines, leads to the formation of the jet and Alfvenic waves. The example shown in this work (Figure~\ref{fig:init}) is this second case. In both cases, the formation mechanism causes a magneto-acoustic shock, which in turn produces chromospheric jets, reaching velocities of $50-100$~km~s$^{-1}$. These shocks pass through the chromosphere in only a few tens of seconds. The chromospheric signature of these shocks is found as short lived RBEs in the line wings of e.g. H$\alpha$ and Ca~{\sc ii} \citep{Rouppe-van-der-Voort:2009ul}. In addition to magnetoacoustic shocks, the release of magnetic tension and mode conversion drive high-frequency transversal waves that propagate into the corona. Within a timespan of a few minutes, cool chromospheric ejected material is heated by the dissipation of currents via ambipolar diffusion to TR temperatures, in broad agreement with the observations of \citet{Pereira:2014eu} and \citet{Skogsrud:2015qq}. The coronal impact of the spicules is the same for both processes that drive the spicules.

\section{Results}~\label{sec:res}

An obvious outstanding question to the study of \citet{Martinez-Sykora:2017sci} addressed in this paper is: What is the impact of type II spicules on the corona?  Before describing our results, we list the TR and coronal synthetic observables of the simulated type II spicules to see whether they fulfill the constraints derived from observations, such as those listed in the introduction \citep[see also][]{De-Pontieu:2017pcd}. For this, we calculate synthetic TR (\ca) and coronal spectral profiles (\fex, \fet\ and \fef), assuming ionization in statistical equilibrium, the optically thin approximation as in \citet{Hansteen:2010uq}  and coronal abundances \citep{Schmelz:2012gf}. These profiles have been integrated along the vertical axis, i.e., as seen from above. Additionally, 
we have degraded the synthetic profiles to the \hinode/EIS spatial (1" per pixel) and spectral resolution ($\sim25$~km~s$^{-1}$) (Figure~\ref{fig:rb}). 

\subsection{Observables and constraints}~\label{sec:obs}

\begin{figure*}[tbh]
\begin{center}
\includegraphics[width=0.95\hsize]{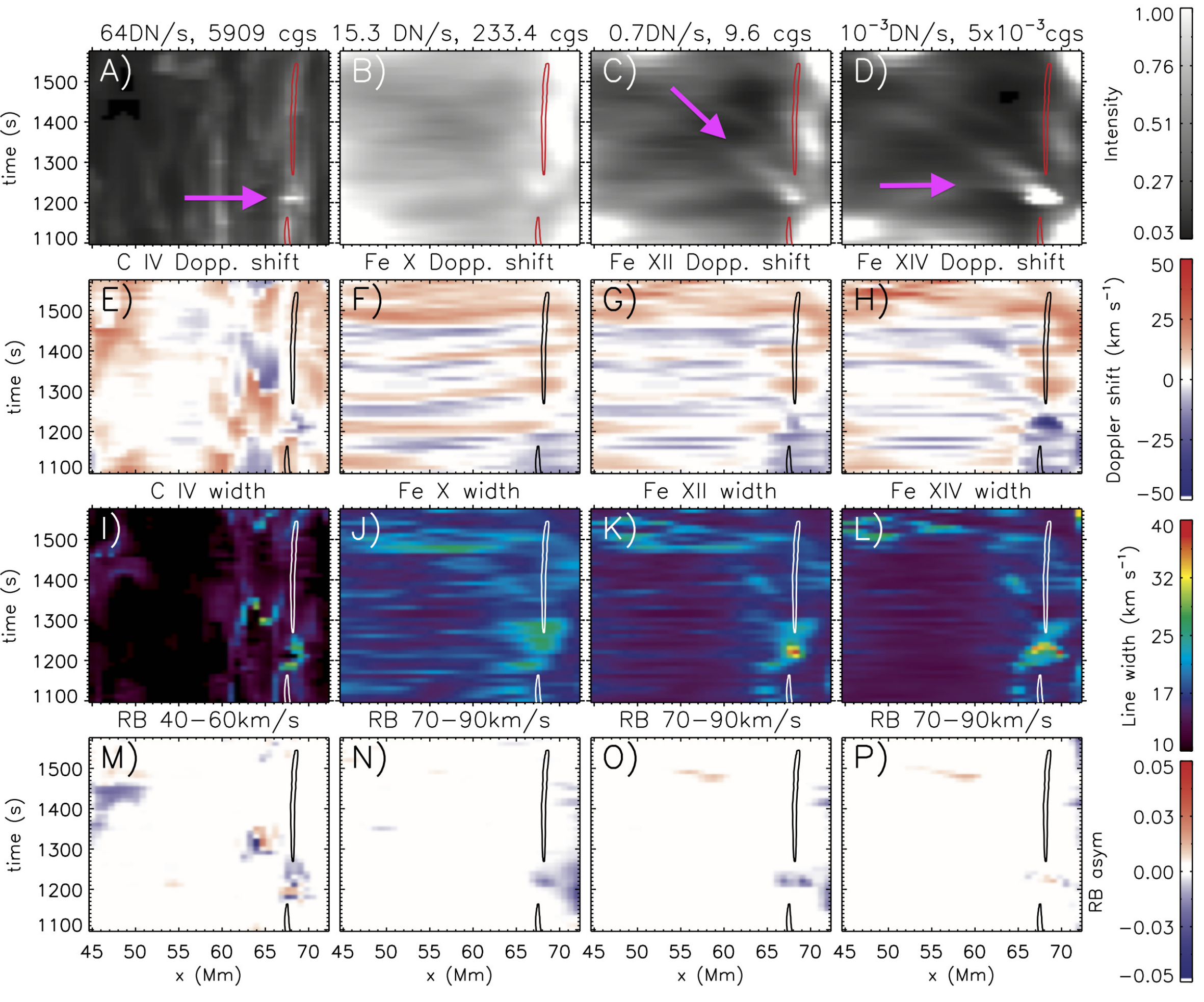}
\caption{Maps of the synthetic intensity (top row), Doppler shift (second row), line width 
(third row) and RB asymmetry (bottom row) for \car\ (first column), 
\fern\ (second column), \fert\ (third column), and \ferf\ (last
column), as a function of space and time reveal coronal signatures associated with a simulated type II spicule (starting at $t=1210$~s and located at $x=68$~Mm, arrow in Panel A). The numbers in the titles on the top row are normalizing factors. The spectral profiles have been degraded to \soho/SUMER (for \car) and to \hinode/EIS (for the iron lines) spectral and spatial resolution. Temperature contours at $10^5$~K at $z=5$~Mm
shows the location of chromospheric spicules penetrating into the corona. The spicule that we focus in this paper corresponds to the contour that goes from $t=1260$~s to $t=1550$~s (Figure~\ref{fig:init}).}
\label{fig:rb}
\end{center}
\end{figure*}

Many observables are associated with the type II spicules in all of the synthesized 
TR and coronal spectral lines. This is illustrated with Figure~\ref{fig:rb} which shows the total line intensity, the Doppler shift, the line width and the RB-asymmetry in the four rows from top to bottom. 

At the beginning of the chromospheric ejection, all four synthesized spectral lines 
show a brightening in intensity (top row, $x=68$~Mm, $t=1210$~s). In the particular case shown here, the synthetic \ferf\  211~\AA\  intensity is too faint to be observed 
with \hinode{/EIS}, i.e., only 
 $ \sim 5\times 10^{-3}$~DN~s$^{-1}$ 
(panel D).  On the other hand, \car\ has enough intensity to be observed with \soho/SUMER and \ferx\ 184~\AA\  and \fert\ 195~\AA\  with \hinode{/EIS} as long as the observations are 
integrated over exposure times of $\sim$ 1, 5, and $>$ 30 seconds or more, respectively (panels A-C).  The various simulated spicules provide a range of intensities at the different wavelengths, and in some cases we find stronger \fert\
intensities (up to 300 erg~cm$^{-2}$~s$^{-1}$~sr$^{-1}$) and weaker \ferx\ than the example shown here. These values are comparable to those in QS by \citet{Brooks:2009bh, Mariska:2013cr}
	
Looking at the intensity observables in the coronal lines (\fert\ and \ferf, panels C and D in Figure~\ref{fig:rb}), one can distinguish two different brightenings, or PCDs, propagating in time and associated with the 
spicules. One of them is a  ``slow" PCD, which at $t=1210$~s is at $x=68$~Mm and 
at $t=1300$~s is at $x=60$~Mm (see arrow in Panel C). The other one is faster, and appears as an almost horizontal intensity brightening in \fert\ and \ferf\ at $t=1230$~s in the region $x=[58,68]$~Mm (arrow in panel D). As described in detail in Sections~\ref{sec:flow} and~\ref{sec:heat}, these observables are a consequence of a combination of a magneto-acoustic shock, (real) flows, and heating processes. Further details of the synthetic PCD observables are discussed and used to interpret \sdo/AIA observations in \citet{De-Pontieu:2017pcd}. 

The synthetic coronal and TR Doppler shifts show only minimal changes (second row) in the spicule location, with the exception of \car\ and \ferf\ which show blue-shifted profiles at the start of the chromospheric ejection. The lack of Doppler shift variation for \ferx\ or very weak Doppler shift variation for \fert\ are due to the line-of-sight integration and also partially due to the relatively low spatial resolution of the \hinode/EIS instrument. Consequently, there is more background emission in the corona along the line-of-sight so that the Doppler shifts are dominated by the background. In this particular simulated spicule most of the \fef\ synthetic intensity (Figure~\ref{fig:rb}) comes from the spicule and very little, if any, is from the overlying and surrounding plasma not associated with the spicules due to the fairly low coronal temperature in this model. Consequently, the synthetic Doppler shift is derived for the most part from the spicular velocities and does not have other possible contributions. 
Similarly to the synthetic \sir\ shown in \citet{Martinez-Sykora:2017sci} the spicule provides a very 
large contribution to the synthetic \c\ intensity. This despite the LOS integration and
pixel resolution. Therefore, we find a large signal in the \car\ spectral lines (and in general for all TR spectral lines), in both 
Doppler shift and Red-Blue (RB) asymmetries. In the bottom row of Figure~\ref{fig:rb} we present the RB asymmetries calculated following the same approach as used by \citet{De-Pontieu:2009fk}  which interpolates in the wavelength axis using spline. We tested the \citet{Klimchuk:2016nx} interpolation method (not shown here) and the calculated RB asymmetries do not differ from the one using spline interpolation shown in Figure~\ref{fig:rb}.

All the spectral profiles become up to 2 times broader compared to their surroundings at the location of spicule acceleration (third row). Similarly observations also show non-thermal broadening in spicules of EUV spectral lines	 \citep[e.g.,][]{McIntosh:2009lr}. 

Despite the low resolution of \hinode/EIS, one can still capture small scale flows by applying RB asymmetry analysis, \citep[e.g.,][]{McIntosh:2009lr,De-Pontieu:2009fk} 
or double Gaussian fitting \citep{Peter:2010fk}. As for \ferf, RB asymmetries  in Figure~\ref{fig:rb}, for this particular case, are 
very small since most of the emission is very localized within the associated spicular plasma.  The synthetic \ferx\ and \fert\ spectral profiles reveal a blue-shifted secondary component at the location of the type II spicule and, for the first time, the synthetic RB asymmetries $\sim 5\%$ are roughly in agreement with  observations \citep[e.g.,][]{De-Pontieu:2009fk,Peter:2010fk}. Indeed, previous models did not reproduce sufficient RB asymmetries \citep[$\sim 0.1\%$][]{Martinez-Sykora:2011oq} due to the lack of type II spicules. 
These RB asymmetries are a clear signal of mass and energy injected into the corona due the combination of a magneto-acoustic shock and real flows, as detailed in Section~\ref{sec:flow}. 

The second spectral component in the TR line (\car, panel M in Figure~\ref{fig:rb}) is found at lower velocities [40-60~km~s$^{-1}$] as compared to the coronal lines [70-90~km~s$^{-1}$] (\fern, and \fert, panels N and O). 
This is due to the acceleration of the plasma as it expands into the corona. This synthetic observational property and the measured speeds are in agreement with observations \citep{McIntosh:2009yf}.  We took into account spatial resolution and  instrumental broadening of the spectral lines since they impact the intensity and location of the secondary component \citep{Martinez-Sykora:2011oq}. 

Could spicules provide enough coronal emission to give any 
signal in the coronal \sdo/AIA channels? The synthetic \fen, \fetsdo, and \fef\ 
intensities associated with the spicule are roughly 
600, 6 and $ 10^{-1}$ \sdo/AIA DN~s$^{-1}$, respectively. 
Consequently, the simulated spicule provides enough synthetic 
intensity contribution from \fen\ to be 
easily observed in the 171 channel. 
For the 193 \sdo/AIA channel, the 
\fetsdo\ intensity coming from the spicule is 
much smaller than from \fern. According to these results, it is very unlikely  
to see any \fert\ contribution coming from the spicule 
in the 193 \sdo/AIA channel  where a typical 
coronal hole and quiet Sun produce something like 9 and 100 
DN~s$^{-1}$, respectively. Despite this, running differences should be able to reveal the intensity 
variations coming from \fetsdo. This is in accordance with 
\cite{De-Pontieu:2011lr} conclusions. Finally, synthetic \fef\ intensity 
coming from the spicule (and any in this simulation) is too small to provide an appreciable signal in \sdo/AIA channel 211 \AA. However, this model has a cool corona for an active region and is limited to a very specific field configuration, note that other more active regions reveal RB-asymmetries in hot lines such as \ion{Fe}{14} \citep{De-Pontieu:2009fk}. 

\subsubsection{Loop width}~\label{sec:width}

Recent observations with Hi-C \citep{Kobayashi:2014sf} suggest that we may resolve the width 
of the many the coronal loops \citep[$\sim500-1500$~km][]{Peter:2013ai,Aschwanden:2017nr}), which are comparable to photospheric granulation scales ($\sim500$~km) \citep{Abramenko:2012qe}. This, according to 
\citet{Aschwanden:2017nr},
suggests that {\it the heating mechanism in the corona is driven by macro-scales
instead of unresolved micro-physics}. 
In addition, observed spicules also seem to have this typical width. 

\begin{figure*}[tbh]
	\begin{center}
		\includegraphics[width=0.95\hsize]{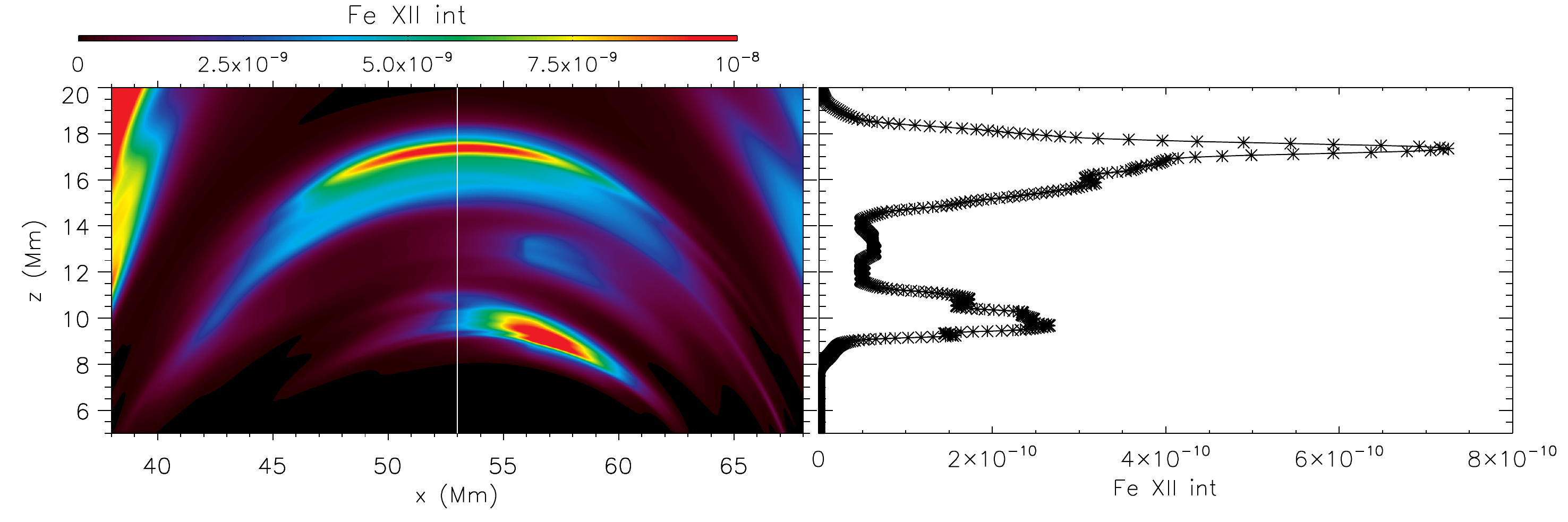}
		\caption{Loops are formed and their
		width is linked to the driving mechanism of the type II spicules. Left panel
	shows the \ion{Fe}{12} emissivity map at t = 1260~s. A vertical cut of \ion{Fe}{12} emissivity  at the apex of the loops (white vertical line in the left panel) is shown in the right panel. Asterisks indicate grid cell positions. }
		\label{fig:wdt}
	\end{center}
\end{figure*}

Figure~\ref{fig:wdt} shows 
a map of the \ion{Fe}{12} emissivity (left) and a vertical cut through a loop (right). The right panel shows that
the width of the loops\footnote{In the following, we use the term “loop” to refer to a region of magnetic field lines that have a common energy and flow source.} associated with the spicules is well resolved, with more than 20 grid points across the loop\footnote{Note that the horizontal axis has a smaller spatial resolution ($14$~km) than the vertical axis in the 
	corona ($20-50$~km).}. 
 The typical width of loops associated with the simulated 
spicules ranges between 300 and 900~km. 
The formation of type II spicules is associated with a PCD and the formation 
of coronal loops \citep{De-Pontieu:2017pcd}. The loops associated with the simulated spicules of this paper have similar widths to the observations reported above (of the order of a few hundreds of kilometers). 
As described in the following sections, the width of the loops associated with the simulated spicules
is determined primarily by the driving mechanism that generates flows in deeper layers, the magnetic connectivity, and heating within 
the magnetic field linked with the spicule (detailed in the following sections). The artificial magnetic diffusion or viscosity plays a minor role in the loop width: in our model the typical dissipation and viscous lengths are 5 grid points, whereas the loop contains between 20 to 100 grid points along the vertical axis (as seen in Figure~\ref{fig:wdt}), and even more in the horizontal axis.

\subsection{Flows and shocks}~\label{sec:flow}

The dynamics occurring in the corona are a consequence 
of both magneto-acoustic shocks and flows: the simulated 
type II spicules drive magneto-acoustic shocks and flows, propagating along the magnetic field at speeds of order
100-200~km~s$^{-1}$, due to the release of magnetic tension, in a manner similar to the whiplash effect, or due to a wave mode conversion driven by photospheric motions. 
In principle, if spicules are driven only by acoustic shocks, they must satisfy the Rankine-Hugoniot (RH) conditions 
which, in a frame of reference moving with the shock front, are given by:

\begin{eqnarray}
\frac{\rho_{post}}{\rho_{pre}} = \frac{(\gamma +1) M^2_{pre}}{2+(\gamma-1)M^2_{pre}} \label{eq:densjump} \\
\frac{u_{post}}{u_{pre}} = \frac{2+(\gamma-1)M^2_{pre}}{(\gamma +1) M^2_{pre}} \label{eq:veljump} \\
\frac{e_{post}}{e_{pre}} = \frac{2\gamma M^2_{pre} - (\gamma -1)}{(\gamma +1)}
\end{eqnarray}

\noindent
where $\rho$, $u$, $e$ and $\gamma$ have their usual meaning. The subscripts ${pre}$ and  ${post}$ refer to pre- and post-shock plasma, respectively \citep{Priest:1982qy}. $M$ is the Mach number ($M=u/c_s$) and the sound speed is $c_s^2 = \gamma P / \rho$. Note that these equations assume no entropy sources outside the shock discontinuity as well as including only the pressure gradient force, i.e., a hydrodynamic fluid without external forces, or the Lorentz force. 

\begin{figure}[tbh]
	\begin{center}
		\includegraphics[width=0.95\hsize]{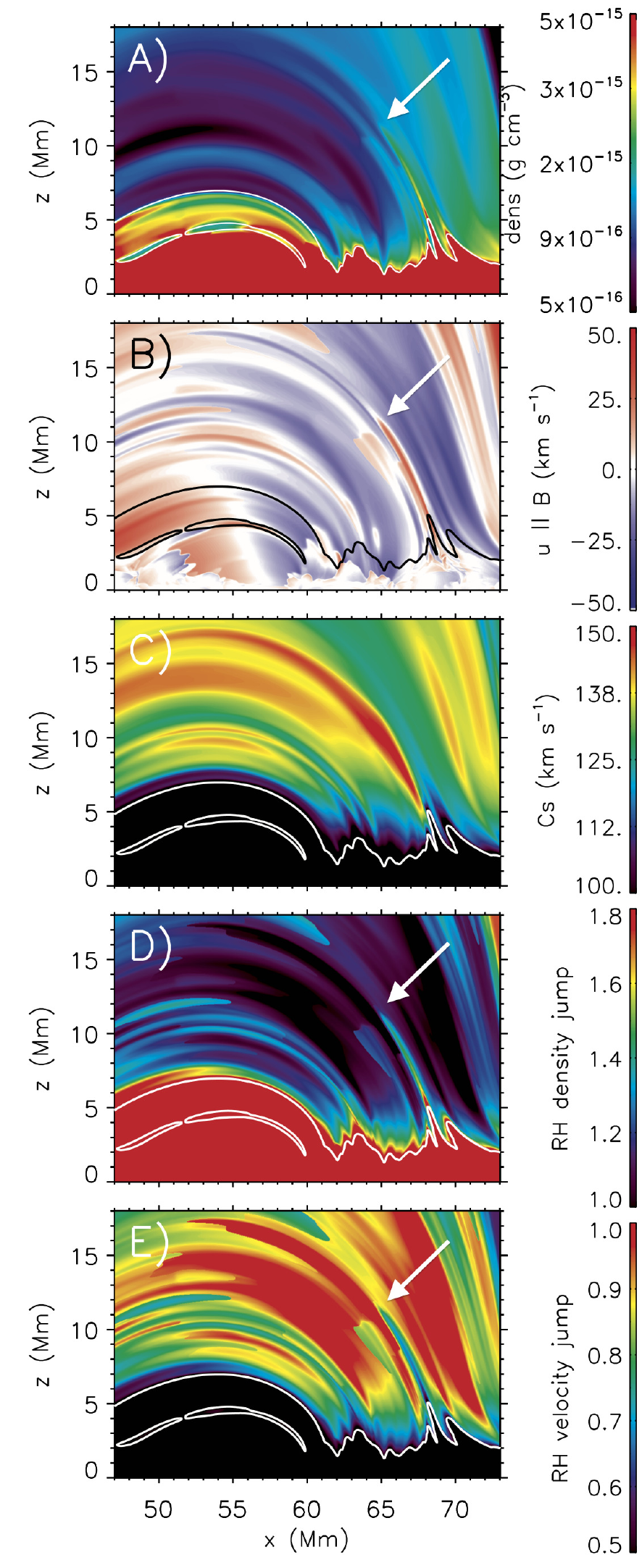}
		\caption{Type II spicules are associated with magneto-acoustic 
			shocks passing through the corona (see arrows). 
			From top to bottom: density, parallel velocity (along the magnetic
			field lines), sound speed,  and maps of the estimated RH density and velocity 
			jumps at $t=1250$s (right hand side of Eqs.~\ref{eq:densjump}-\ref{eq:veljump}, respectively). The solid contour in all panels represents a temperature of $10^5$~K. The lowest part of the spicule is situated at $x\sim 68$~Mm.} 
		\label{fig:shock}
	\end{center}
\end{figure}

The example shown in Figures~\ref{fig:init} and~\ref{fig:rb} is examined in detail in Figure~\ref{fig:shock}, where panel A shows the density stratification and with a density jump located at [x,z]=[65,11]~Mm (see arrows). The jump is easier to discern in the parallel velocity map (along the magnetic field) shown in panel B. 
Taking into account the coronal sound speed ($\sim150$~km~s$^{-1}$, panel C) and the fact that the shock travels through the corona at 150~km~s$^{-1}$, the RH relations (right hand side of Equations~\ref{eq:densjump}-\ref{eq:veljump}) can be calculated from the velocity along the magnetic field (panel B) subtracting the front shock speed (150~km~s$^{-1}$), sound speed (panel C) and the adiabatic parameter $\gamma$ (1.66) shown in panels D and E.  From there, one would predict a density jump of 1.4 and velocity ratio of 0.6 (panels D and E). However, the jumps in the corona are of the order of 1.8 and 0.75 for the density and velocity, respectively (panels A and B).
This disagreement is due to the fact that these jumps depend not only on the gradient of pressure but also on the Lorentz force, which is absent in 1D hydro-models, and other entropy sources working on the bulk flow, in particular some of the plasma moving with the shock wave is heated via Joule heating and/or thermal conduction and thus becomes part of the corona.
Note that this perturbation (or jump) will travel 1~Mm in 10~seconds, this explains: 1) the short-lived brightening in the AIA channels due the increase in density; 2) the non-thermal line broadening and RB asymmetries in the TR and coronal spectral lines at the beginning of the spicule formation are due to the velocity jump and the difference in velocity between the loop associated with the spicule and its surroundings; 3) finally, this perturbation, which travels along the loop, leads to the synthetic ``slow" PCD.

\subsection{Mass and energy deposition in association to spicules}~\label{sec:heat}

\begin{figure}[tbh]
	\begin{center}
		\includegraphics[width=0.95\hsize]{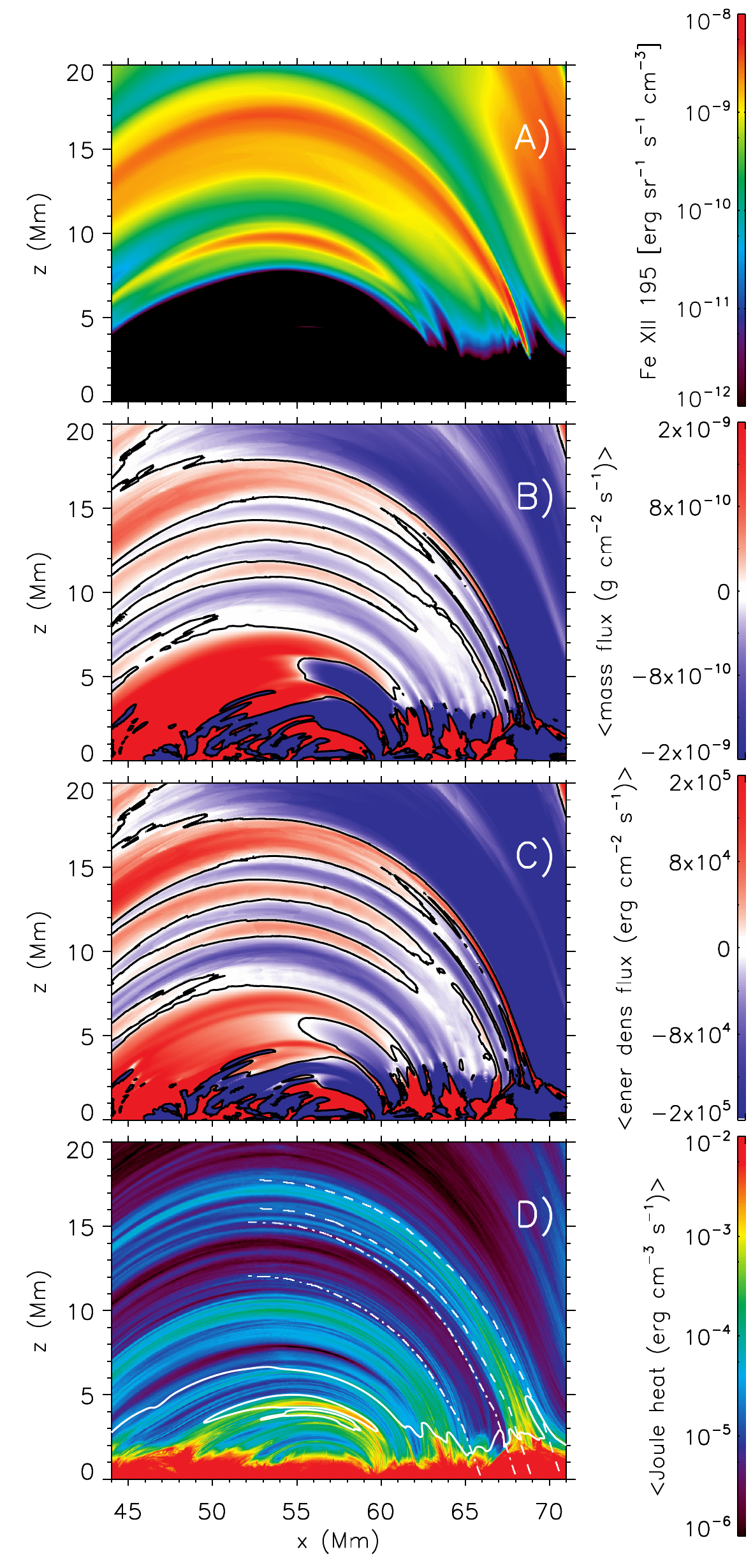}
		\caption{Type II spicules provide mass and energy density flux as well as heating to the upper atmosphere, thereby heating and filling ``individual" strands. From top 
			to bottom: \fert\ emission, mean mass and energy flux along the magnetic field and Joule 
			heating integrated over the 
			lifetime of the spicule. Solid white and black contours correspond to a temperature of $10^5$~K and zero value, respectively. Note how the heating is mostly concentrated in magnetic field lines  
			associated with the spicule compared to the surroundings. The dashed lines in panel D outline the limits used to calculate the average heating shown in Figure~\ref{fig:heat}. See corresponding Movie~1.} 
		\label{fig:pcd}
	\end{center}
\end{figure}

The mass and energy deposition in the corona from a spicule results from a combination of shocks, flows, Joule heating, and thermal conduction. The plasma is heated along the loop through Joule heating (see below) resulting from numerical diffusivity. 
The artificial diffusion will not impact the amount of the Joule heating dissipation  \citep{Galsgaard:1996lf,Gudiksen:2005lr,Gudiksen:2005fk,Hansteen:2007dt,Peter+Gudiksen+Nordlund2004, abbett2000,Rempel:2017zl}. In addition, the diffusion length scales are smaller than the loop width so will not change the latter.
 Density and energy jumps become less sharp in the corona 
due to thermal conduction (panel A, Figure~\ref{fig:shock}), in contrast to the velocity jump (panel B). Therefore, the synthetic ``slow" PCD intensity becomes less sharp and weaker as the perturbation propagates through the corona. In addition, flows provide mass and energy flux into the corona which remain hot and do not return to chromospheric conditions due to strong Joule heating and thermal conduction. Hence, loops associated with spicules (Panel A of Figure~\ref{fig:pcd}) are filled with mass and energy by the flows, as shown in panels B and C of Figure~\ref{fig:pcd} and Movie~1 and currents shown in panel D. Both the figure and movie are made by integrating in time, the former over the lifetime of the spicule and the latter integrating forward in time from the initial formation of the spicule to the current frame, i.e., the time range shown in the top panel. This figure and movie reveals that flows and currents are not always located exactly along the same magnetic field lines. Instead, currents and flows occur within the bulk of magnetic field that drives the formation of the spicule. 
Prior to the formation of the spicule (A) we can see the remnants of a decaying hot loop. This hot loop was formed in association with a different spicule (B) that occurred earlier and at the conjugate footpoints of the field lines of spicule A. By the time of the formation of spicule A, the previous spicule had already vanished.  The impact of spicule B and its associated heating is negligible in our calculations.

Let us quantify the mass and energy deposition into the corona. 
For this, first, we focus on the various fluxes due to advection. 
Averaged over the lifetime of the spicule (10 minutes), we find a net positive mass flux ($\sim 10^{-9}$~g~cm$^{-2}$~s$^{-1}$) and energy flux due to advection, ($\sim 10^{5}$~erg~cm$^{-2}$~s$^{-1}$) for this loop associated with the spicule (red loop located at $(x,z)\approx(57,15)$~Mm in panels B and C of Figure~\ref{fig:pcd} and Movie 1). 
These fluxes are measured as they cross the transition region ($T=10^{5}$~K).
At first glance, these numbers seem lower than the observational estimates of energy flux by an order of magnitude \citep[$\sim 2\times 10^{6}$~erg~cm$^{-2}$~s$^{-1}$, ][]{De-Pontieu:2011lr}. Note that these numbers have been used by, e.g., \citet{Klimchuk:2012kx,Klimchuk:2014fk}. In other words, assuming a spicule diameter of 300~km, the energy deposition due to advection in this simulation is $\sim 9\times 10^{19}$~erg~s$^{-1}$. The discrepancy with estimates from observations is due to the following: 1) 
those calculations are based on assumptions driven by estimates from the observations such as constant velocity in time ($100$~km~s$^{-1}$, over 2-4 minutes), 
in contrast to the simulated spicule's velocity which varies within $\sim\pm150$~km~s$^{-1}$ over the spicule lifetime. 2) We integrated over 10 minutes, i.e., the spicule's lifetime, whereas \citet{De-Pontieu:2011lr} integrate between 2-4 minutes, using the lifetime of the brightening\footnote{Note that at that time the full lifetime of spicules was unclear \citep{Pereira:2014eu}.} or even a shorter period of time by \citet{Klimchuk:2012kx,Klimchuk:2014fk}, which gives even less total heating. 3) Last and crucially, our calculations above, including only advective terms ignore a critical contribution to spicule energetics: the Joule heating including the dissipation of transversal wave energy. 

\begin{figure}[tbh]
	\begin{center}
		\includegraphics[width=0.95\hsize]{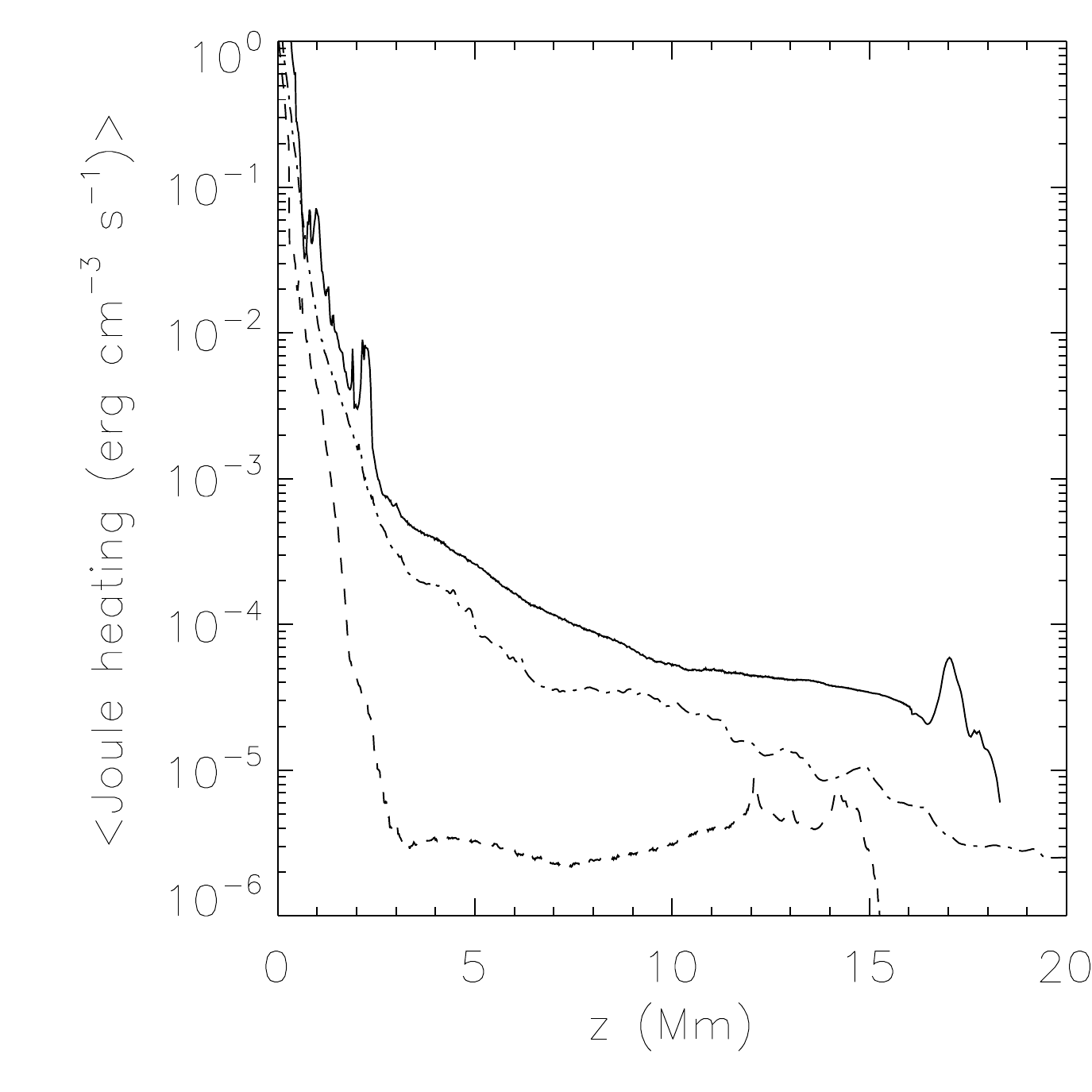}
		\caption{Mean Joule heating integrated over 10 minutes and width 
			as a function of height for the loop associated with the spicule ($\sim 1.5$~Mm wide, outlined by the dashed lines in Panel D Figure~\ref{fig:pcd}) and without spicules ($\sim 2.5$~Mm wide, outlined by the dot-dashed lines in Panel D Figure~\ref{fig:pcd}) are shown with solid and dashed lines respectively. For comparison, we included the mean Joule heating as a function of height for the whole simulated domain of the non-GOL simulation, i.e., without ambipolar diffusion (dot-dashed line).} 
		\label{fig:heat}
	\end{center}
\end{figure}

The simulated spicules reveal the importance of the contribution coming from  Joule heating ($3\times 10^{-5}-10^{-2}$~erg~cm$^{-3}$~s$^{-1}$) shown in the bottom panel of Figure~\ref{fig:pcd} and Figure~\ref{fig:heat}.  Assuming that the spicule is within $z=[3,8]$~Mm and a radius of 300~km, the energy deposition due to  Joule heating is $\sim 2.4\times 10^{22}$~erg~s$^{-1}$. However, the Joule heating associated with the spicule is not
confined to the cold plasma of the spicule. Currents extend and propagate along the magnetic field associated with the spicule and surroundings \citep[compare with ][who assumed that all the energy is located inside the spicule and ignored any current propagating into the corona]{Klimchuk:2012kx}. Taking all of this into account: the region associated with the spicule where the current plays a role (see dashed lines in Figure~\ref{fig:heat} $\sim1.5$~Mm wide and $z\geq3$~Mm along the associated loop), the total deposited energy due to Joule heating is  $\sim 1.2\times 10^{24}$~erg~s$^{-1}$, i.e. up to four orders of magnitude greater than that associated with the advective terms. Consequently, the dominant contribution to spicule heating comes from Joule heating instead of the energy flux due to kinetic and enthalpy fluxes. 
Taking into account the Joule heating and energy flux due to the advection of the plasma we find that this is enough to maintain the loops associated with the spicules heated to coronal temperatures for several tens of minutes. 

The numbers provided above are valid for a single spicule in the model. Other spicules show similarities but we find that the contributions from the various sources to coronal heat input can vary widely. In fact, in the model, some spicules are associated with strong heating and rather small amplitude flows and small shocks, while others have strong advection and small Joule heating. Thus, Joule heating, energy flux due to advection, and wave energy can all dominate the energy budget of individual spicules.

We find that the fast PCD in the coronal spectral lines propagates at Alfv\'enic speeds and is a direct consequence of Joule heating. In fact, both PCDs are associated with magnetic energy deposition in the corona, driven by the type II spicules. At the very beginning of the spicule formation, currents (and Joule heating) form along the magnetic field lines (panel D, Figure~\ref{fig:pcd} and Movie~1). These currents are driven by the release of the magnetic tension in the chromosphere, which leads to transverse waves as well as electric current propagating along the magnetic field at Alfv\'enic speeds. Note that transverse waves produce currents. Magnetic energy is dissipated and heats the plasma within the loop in our model by artificial (numerical) diffusion. On the Sun, one may expect a cascade to even smaller scales before dissipation sets in, e.g., via resonant absorption \citep{Okamoto:2015fr} of transverse waves or via the Kelvin-Helmholtz instability \citep{Antolin:2017}. Although not uniform in space nor constant in time, the current is generated continuously over the lifetime of the spicule and continues to exist even after the spicule has fallen back to the chromosphere.  
In addition, currents do not remain in exactly the same location (or on a single magnetic field line) but `move' to those magnetic field lines that are associated with the spicule (Movie~1), leading to currents `meandering' through the coronal volume within the loop associated with the spicule (see next section for details on the width of the spicules and associated loops). 

It is important to point out that the coronal Joule heating outside the loop structure associated with the spicule is, at least, an order of magnitude less than the Joule heating given by the formation of the type II spicule. 
This can be seen in both the bottom panel of Figure~\ref{fig:pcd} and in Figure~\ref{fig:heat}. 
In the bottom panel of Figure~\ref{fig:pcd}, the Joule heating inside the loop (within the white dashed lines) seen in light-blue is more than an order of magnitude greater than its surroundings (e.g., see the region within the dot-dashed lines) seen with dark blue or black color. One can see a second loop structure which also has large Joule heating at roughly  at $x=64$~Mm associated with another type II spicule. 
Figure~\ref{fig:heat} shows the Joule heating as a function of height, averaged over 10 minutes and spatially limited to the loop associated with the spicule in solid line ($\sim 1.5$~Mm wide, outlined by the dashed lines in panel D of Figure~\ref{fig:pcd}). We also show (dashed line) the Joule heating in a region that is not associated with any type II spicule ($\sim 2.5$~Mm wide, outlined by the dot-dashed lines in panel D of Figure~\ref{fig:pcd}). For comparison, we also show Joule heating as a function of height of the whole box for the non-GOL simulation (dot-dashed line). Note that for regions without spicules, the Joule heating is not only lower but remains mostly constant as a function of height in the corona. We note that the average Joule heating in the non-GOL simulation is smaller than in the loops associated with the spicules (in the GOL simulation) but higher than in regions without spicules (in the GOL simulation). This is due the fact that the ambipolar diffusion reduces the amount of current that reaches into the corona in regions without spicules.
 It would be of great interest to expand this simulation into 3D in order to have a better description of the magneto-convective motions. Note that, in general, braiding is more efficient in 3D models as compared to in 2D models and the formation of type II spicules depends on the magneto-convective motion of the magnetic field lines in the chromosphere, as pointed out in \citet{Martinez-Sykora:2017sci} the convective motions move the magnetic field, accumulates it at the surface and increase its tension.
 
\begin{figure}[tbh]
	\begin{center}
		\includegraphics[width=0.95\hsize]{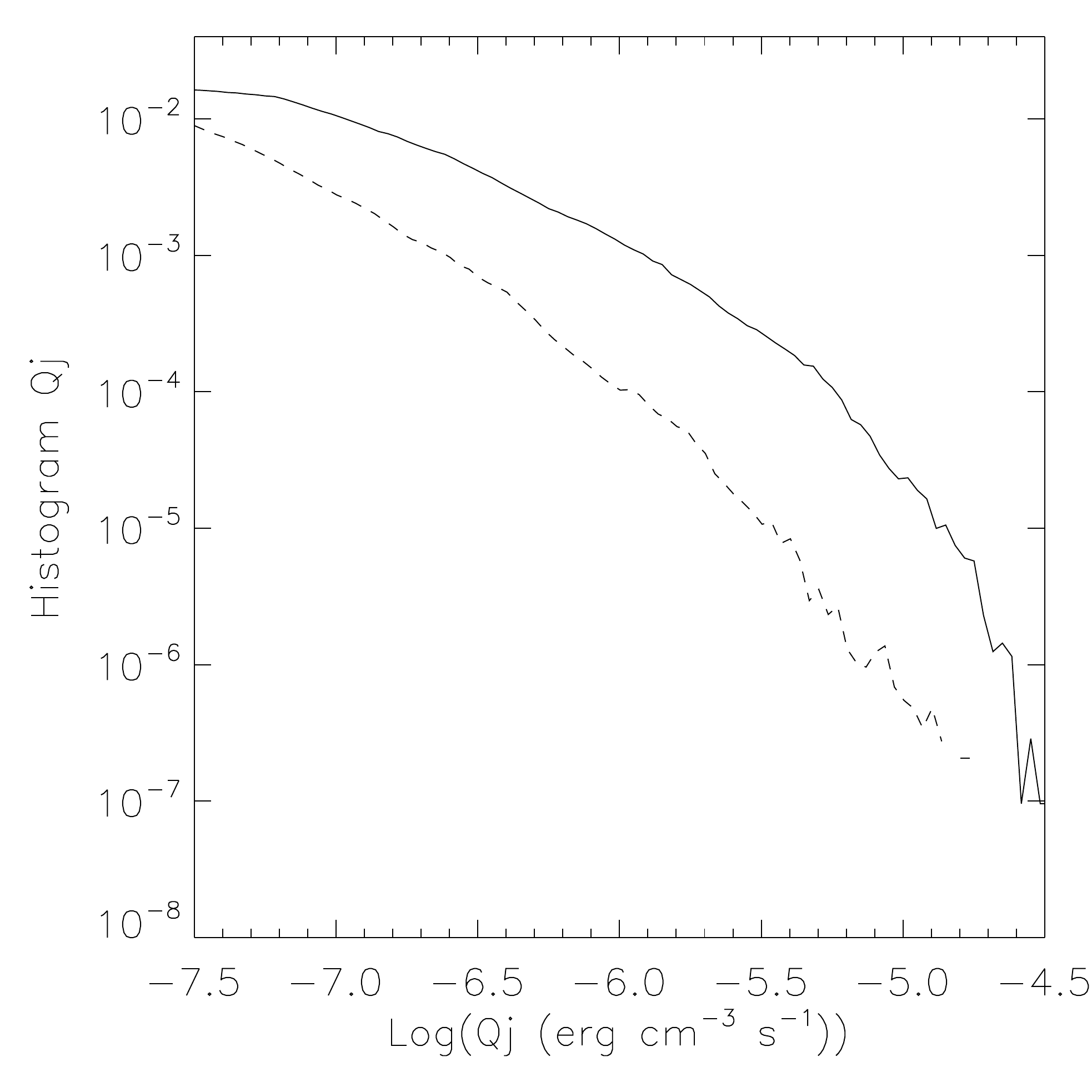}
		\caption{The Joule heating histogram in the corona in the loop associated with the spicule (within the dashed lines in panel  D in Figure~\ref{fig:pcd}) shown with solid line reaches greater values and follows a different profile than in coronal regions not affected by spicules shown with dashed line (within the dot-dashed lines in panel D Figure~\ref{fig:pcd}).} 
		\label{fig:histo}
	\end{center}
\end{figure}

Another way to visualize the heating distribution in loops associated with spicules and in loops not associated with spicules is by constructing Joule heating histograms within these two set of loops, as shown in Figure~\ref{fig:histo}. 
The loop associated with the spicule not only has locations with more energy but also a different slope (solid line) in comparison with the loops without type II spicules. One must be careful in assuming that the Joule heating histogram corresponds to Joule heating events \citep[usually used in observations][]{Benz:2002le} since in these models it is very difficult to isolate the events \citep{Guerreiro:2015jk,Hansteen:2015qv,Guerreiro:2017bq}. In addition, one must interpret these results qualitatively instead of quantitatively since the simulation is only 2D and the magneto-convective motion and the nature of reconnection events can change when expanding into 3D. 

\subsubsection{Loop width determined by the processes associated with type II spicule formation}~\label{sec:width2}

The presence of the ambipolar diffusion and the generation of type II spicules
tend to confine the propagating currents and waves to magnetic field lines associated with the spicule as shown in the previous section. 
The simulation without ambipolar diffusion (non-GOL) also generates transverse waves and currents due to
the dynamics in the lower chromosphere. However, these currents are generated and 
propagate in non-specific locations due to the magnetic convection. Consequently, the 
non-GOL simulation shows currents and Joule heating in extended regions in the corona, as 
seen in Figure~\ref{fig:nongol} where many thin light-blue loops are evident in the corona \cite[][compares the global aspects of Joule heating between the non-GOL and GOL simulations]{Martinez-Sykora:2017gol}. 
In contrast, the ambipolar diffusion confines currents to regions associated 
with the generation of type II spicules. 
The propagation of transverse waves and currents into the corona is limited to a specific set of coronal magnetic field lines as shown with light blue color in panel D of Figure~\ref{fig:pcd}. In contrast, other regions have much less Joule heating (dark blue) than those field lines connected to  type II spicules 
as mentioned in the previous section. This leads to greater thermal contrast in the corona in the GOL simulation as compared to the non-GOL simulation, and, hence, a formation of EUV coronal loops associated with the spicules.

Similarly, the ejected mass and energy into the corona from the spicule
is confined  within the loop associated with the spicule (see Movie 1). 
Since the expansion of the magnetic field mostly occurs between the photosphere
and middle chromosphere, in the upper chromosphere and corona the
magnetic field lines barely expand. Consequently the ejected 
mass and energy into the corona 
have roughly the same width along the loop (see Movie 1). 
This is also clear when viewing the 
synthetic coronal loops shown in Figure~\ref{fig:wdt}. 
As a result of this, the loop formation due to 
the mass deposition is determined by the width of the spicule in the chromosphere.  
Consequently, the flows and currents associated with the simulated spicules are localized 
in loops within 500 to 900~km full width. Flows are slightly 
more confined ($300-600$~km) than the currents. Still, one must be aware of the 2D limitation of these simulations which could have greater expansion.

\begin{figure}[tbh]
	\begin{center}
		\includegraphics[width=0.95\hsize]{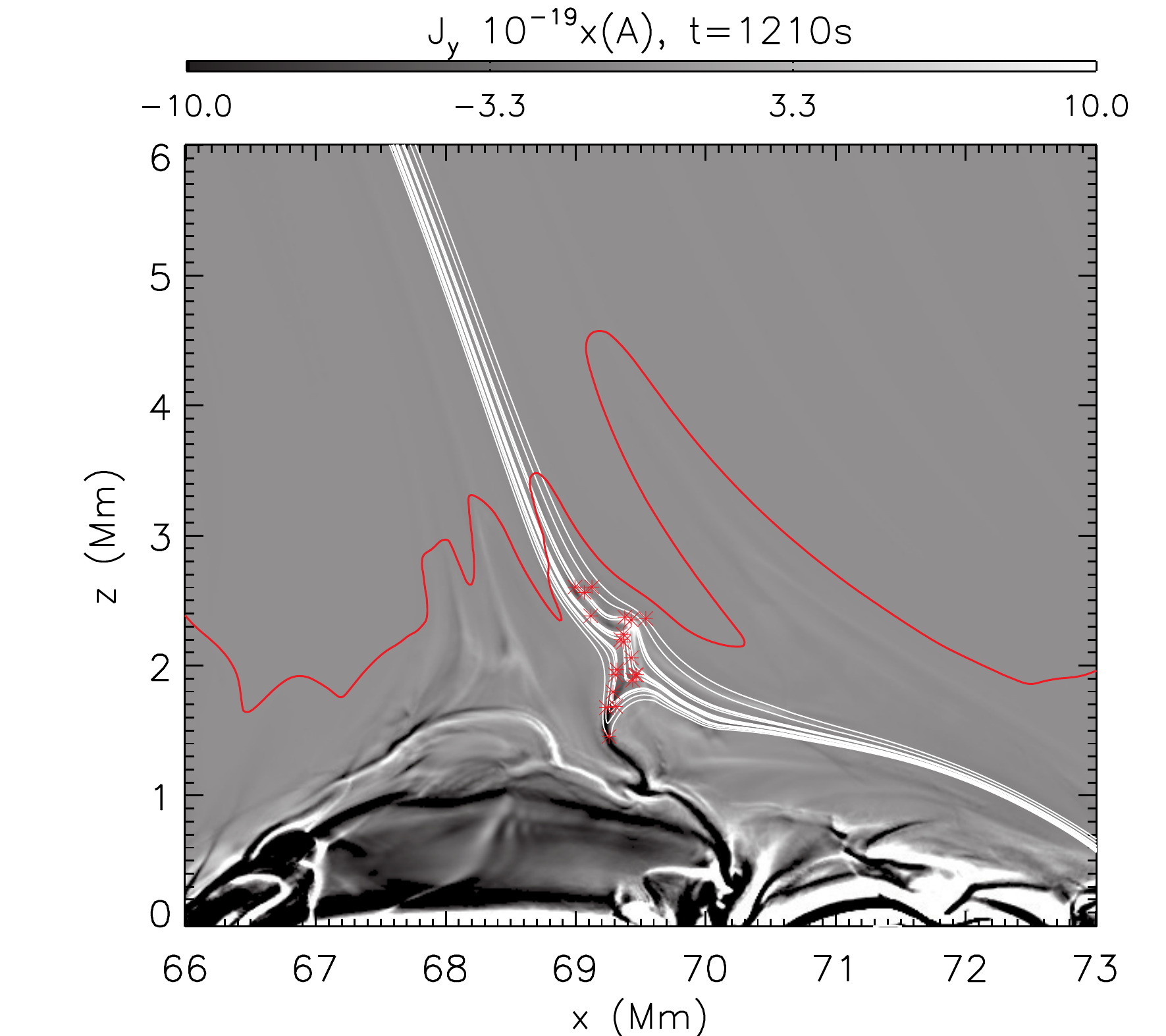}
		\caption{Current perpendicular to the plane allows to identify the length (width) of the driver. The seeds 
		(red asterisks) of the magnetic field lines are located along the current that drives the spicules ($(x,z)\approx(69.5,1.8)$~Mm). The red line corresponds to temperatures of $10^5$~K and allows to locate the spicule ($(x,z)\approx(69,3)$~Mm). See corresponding Movie~2.} 
		\label{fig:scl}
	\end{center}
\end{figure}

The size of the spicule and 
consequently the associated coronal loop is determined by the magnetic topology 
at the formation of the spicule. Figure~\ref{fig:scl} shows the current perpendicular to the plane in an early state of the formation of the spicule. The driver is easy 
to identify and follow in time with the current (Movie 2). In addition, we added magnetic field lines seeded (red asterisks in the Movie~2) along the current structure that drives the spicule, i.e., the seeds do not follow the plasma nor are forced to the same position in time. In the movie one can see that this current is formed (built up) in the photosphere. Its length (typical in photospheric granulation, i.e., $\sim500$~km) is defined due to convective motions and shapes the size of this current structure. Eventually, it penetrates into the chromosphere and drives the spicule where plasma $\beta$ is close to unity. 
Note that the field lines 
are always within the spicule, and the currents traveling into the corona last longer than the movie. During all this time, as shown in Figure~\ref{fig:pcd}, the current traveling into the corona is confined within the loop structure associated with the spicule, in particular to the currents shown in the Figure~\ref{fig:scl} and Movie 2.

\subsection{Various processes throughout the atmosphere}~\label{sec:strat}

In the simulated atmosphere, the Joule heating in the regions associated with the spicule follows three different power-law decays with height as shown in Figure~\ref{fig:heat}. In general the structure of the field just above the photosphere will reflect scales on the order of granulation, while at greater heights it is meso- or super-granular sized field structures that dominate. These different regimes will be reflected in the Joule dissipation of braided field lines. In particular, for the field lines associated with spicules we 
have additional processes that set the scale height of magnetic field dissipation.
The lower chromosphere has the steepest decay with height  ($\sim-1$), then the upper chromosphere ($\sim-1/5$) and finally the least decay is in the corona ($\sim-1/20$). Inside the chromosphere, this property was also observed for the ambipolar velocity as detailed in Figure~7 in \citet{Martinez-Sykora:2017gol}. There the lower and upper 
chromosphere are characterized by two different dominant processes which
lead  to different spatial and time scales. 
In the lower chromosphere, which is dominated by 
magneto-acoustic shocks, the current and Joule heating are smoother, more spatially extended than in the upper chromosphere.
In contrast, in the upper chromosphere and corona, currents driven by spicules are concentrated in narrow regions along the spicules or loops. 
So, which process leads to different decays with height in the corona? In the simulated chromosphere ($z\approx[1,8]$~Mm), the Joule heating is due to ambipolar diffusion whereas \citep{Martinez-Sykora:2017gol}, in the corona, it is mostly due to artificial diffusion. Note that this slope strongly depends on the current that penetrates the corona coming from deeper layers (compare solid and dashed line slopes) and here, the formation of the spicules plays a crucial role. 
Since the model is limited to 2.5D, type II spicules drives only transverse waves which, in a fully 3D model, may be kink or sausage waves (or other modes) and the energy deposition will vary, depending on the type of waves. Details about the transverse wave formation and energy flux are given in \citet{Martinez-Sykora:2017sci}.

\begin{figure}[tbh]
	\begin{center}
		\includegraphics[width=0.95\hsize]{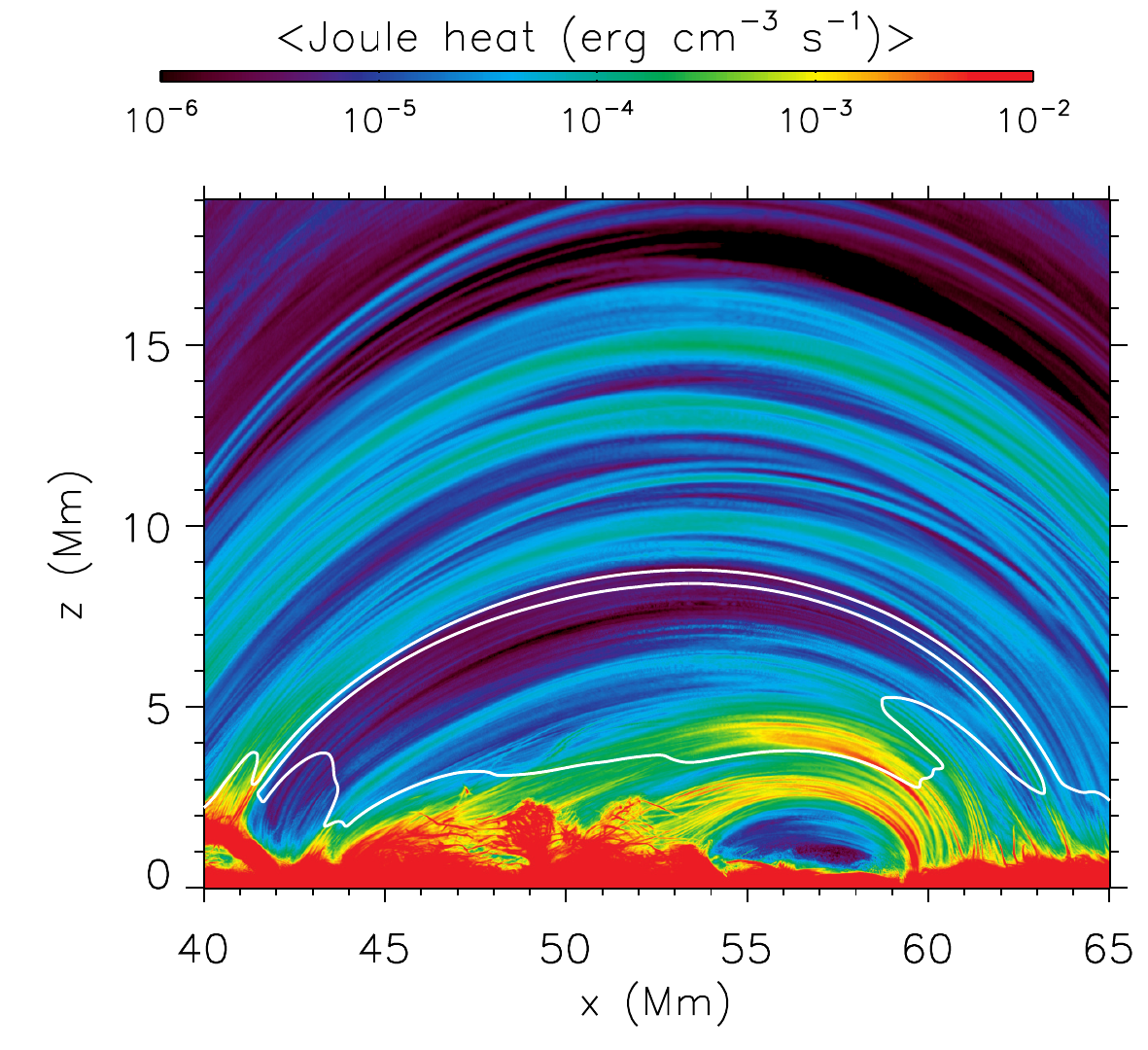}
		\caption{The non-GOL simulation shows more uniform heating distribution in contrast to the GOL simulation. Mean Joule 
			heating integrated over the 
			lifetime of 10 minutes is shown in logarithmic scale for the non-GOL simulation as shown for the GOL simulation in Figure~\ref{fig:pcd}.} 
		\label{fig:nongol}
	\end{center}
\end{figure}

\section{Discussion \& Conclusions}~\label{sec:dis}

We report on the impact of type II spicules in the corona using 2.5D radiative MHD
simulations. Our TR and coronal synthetic spectral profiles provide an 
explanation for the observations 
of so-called RB asymmetries \citep{De-Pontieu:2009fk,Peter:2010fk,Bryans:2010lr} and coronal brightenings in the \sdo/AIA counterpart. Our simulation
is able to reproduce RB asymmetries of the same magnitude as seen in observations. This in contrast to 
previous simulations which failed to reproduce observed RB asymmetries
\citep{Martinez-Sykora:2013ys}. All of these observational diagnostics are found to be signatures of the impact 
of type II spicules on the corona. 

Spicules contribute to the corona through the loops that are associated with the spicules.  Mass, kinetic energy flux, and 
Joule heating, are injected into the corona. Type II spicules are generated by the release of magnetic tension, which drives 
acoustic shocks and flows, and transverse waves and electric currents that travel along the magnetic field. The flows fill the loops with plasma. In addition, the temperature and density jumps are smeared out in the corona due 
to the thermal conduction. The RB asymmetries, line broadening and intensity increase that occur in the early stages of the spicule are a consequence of these processes. Joule heating plays a crucial role in
heating the plasma and maintaining loop temperatures against energy losses in the corona. 
This process leads to synthetic PCDs similar to those observed in coronal EUV imaging. 

Ambipolar diffusion is a key process for the formation of type II spicules and, in addition, to localize the
mass, and energy flux including waves and Joule heating in the loop associated with the spicule. 
All of these processes are localized within a few hundred kilometers, whereas in the surrounding corona, the amount of heating is at least an order of magnitude lower. Even if we compare with the non-GOL simulation, the heating in the simulated corona that lack spicules (in the GOL simulation) is an order of magnitude lower and the variation of Joule heating with height is almost constant, contrary to those loops associated with spicules. Since the amount of current build up in the corona via convective motion is larger in 3D than in 2D, our results must be taken qualitatively rather than purely quantitatively. Similarly, any 3D results without ion-neutral interaction effects must also be interpreted carefully. 

The width of synthetic loops associated with spicules seems in agreement 
with observations \citep{Aschwanden:2017nr}. Consequently, the width of coronal loops, at least for those  formed
in association to spicules, are not governed by micro-physics,
but the macro-physics included in the model. These macro-physics determine
the heating and mass deposition.  The typical width 
of the loops associated with the spicules are set by the magnetic field 
topology structures formed in the photosphere. These structures drive both the currents that heat the associated loops in the corona and flows. Consequently, both flows and heating have roughly similar width and guided by the ambient magnetic field. 

In this model, it is very hard to separate between braiding and wave heating. In addition, note that the braiding in this model changes the classical braiding picture. Most of the magnetic tension in the photosphere and lower chromosphere is generated by convective motion. However, this magnetic energy does not propagate to greater heights easily and mostly accumulates in the lower atmosphere. Spicule formation allows the release and propagation of magnetic energy and tension to greater heights thanks to the ambipolar diffusion. Whereas, in other regions, the ambipolar diffusion dissipates magnetic energy into thermal energy within the chromosphere. As detailed here and in \cite{Martinez-Sykora:2017sci}, the released tension also drives Alfvenic waves. In the Sun, these waves may be either Alfven waves, fast modes, kink waves, sausage modes or similar. It is important to study which modes are being generated since the Alfvenic energy flux conversion and dissipation depend on the type of the wave. 
Note that, since the model is limited to 2D, the magneto-convective motion and consequently, most likely, number of events may change when we expand this into 3D. 

Last, but not least, the formation of type II spicules generates currents that propagate
along the magnetic field at Alfvenic velocities into the corona. These currents are generated 
over the lifetime of the spicule and confined within the location of the spicule's driver but not always on the exact magnetic field lines (see Movie~1) containing the cold spicular material. The amount of energy contained in these currents is enough to sustain the associated coronal loops for several tens of minutes. 
This is in contradiction to previous findings which assumed a much smaller amount of energy release and over-simplified the heating mechanism driving the formation of type II spicules. In our model currents are dissipated by ambipolar diffusion in the chromosphere and TR, and with artificial diffusion in the corona. Since the simulation is able to reproduce many of the observables, the artificial diffusion may mimic the un-resolved physics
that dissipate the sharp currents and turbulence quite well. To resolve this issues one must perform specific studies to address further questions. 

The results presented here and in \citet{Martinez-Sykora:2017sci} 
are of great interest for further studies since they provide constraints for
ad hoc terms that other models use, e.g., episodic heating profiles as a function of height for 
coronal rain simulations such as \citep{Antolin:2010qv,Xia:2017uk}, or wave amplitudes and fluxes for waves studies \citep{Antolin:2015wd}. The heating profile shown in Figure~\ref{fig:heat} could be used as input/driver for many other studies. However, one must take into account that the heating is localized, episodic, and not always at the exact same magnetic field lines (Movie~1, \citet{Hansteen:2015qv,Guerreiro:2015jk}).

It is important to mention that the spicules simulated here do not have a straightforward relation between the amount of mass and energy flux, wave energy, currents and Joule heating.  Indeed, we found strong heating processes associated with spicules with rather low flows and small shocks, or vice versa. 

\section{Acknowledgments}

We gratefully acknowledge support by NASA grants, NNX16AG90G, NNH15ZDA001N, NNX17AD33G, and NNG09FA40C (IRIS), NSF grant AST1714955. This research has received funding from the UK Science and Technology Facilities Council (Consolidated Grant
ST/K000950/1) and the European Union Horizon 2020 research and innovation programme (grant agreement No.
647214). This research was supported  by the 
European Research Council under the European Union's Seventh Framework 
Programme (FP7/2007-2013) / ERC Grant agreement nr. 291058. 
The simulations have been run on clusters from the Notur project, 
and the Pleiades cluster through the computing project s1061, s1472 and s1630 from the High 
End Computing (HEC) division of NASA. We thankfully acknowledge the 
support of the Research Council of Norway 
through grant 230938/F50, through its Center of Excellence scheme, project number 262622,  and through grants of computing time from the 
Programme for Supercomputing. This work has benefited from discussions at 
the International Space Science Institute (ISSI) meetings on 
``Heating of the magnetized chromosphere'' where many aspects of this 
paper were discussed with other colleagues.
To analyze the data we have used IDL.

\bibliographystyle{aa}
\bibliography{aamnemonic,collectionbib}

\end{document}